\documentclass[twocolumn,superscriptaddress,showpacs,preprintnumbers,amsmath,amssymb,prl]{revtex4}

\usepackage{graphicx}
\usepackage{dcolumn}
\usepackage{bm}

\begin{document}

\title{Formation and Stability of Synaptic Receptor Domains}

\author{Christoph A. Haselwandter}
\altaffiliation{Present address: Department of Applied Physics, California
Institute of Technology, Pasadena, CA 91125}
\affiliation{Department of Physics, Massachusetts Institute of Technology, Cambridge, MA 02139}

\author{Martino Calamai}
\affiliation{Biologie Cellulaire de la Synapse N\&P, INSERM UR497, Ecole Normale Sup\'erieure, 46 rue d'Ulm, 75005 Paris, France}

\author{Mehran Kardar}
\affiliation{Department of Physics, Massachusetts Institute of Technology, Cambridge, MA 02139}

\author{Antoine Triller}
\affiliation{Biologie Cellulaire de la Synapse N\&P, INSERM UR497, Ecole Normale Sup\'erieure, 46 rue d'Ulm, 75005 Paris, France}

\author{Rava Azeredo da Silveira}
\affiliation{Department of Physics and Department of Cognitive Studies, Ecole Normale Sup\'erieure, 24 rue Lhomond, 75005 Paris, France}

\date{\today}

\begin{abstract}
Neurotransmitter receptor molecules, concentrated in postsynaptic domains along with scaffold and a number of other molecules, are key regulators of signal transmission across synapses. Employing experiment and
theory, we develop a quantitative description of synaptic receptor domains in terms of a reaction-diffusion model. We show that interactions between only receptor and scaffold molecules, together with the rapid diffusion of receptors on the cell membrane, are sufficient for the formation and stable characteristic size of synaptic receptor domains. Our work reconciles long-term stability of synaptic receptor domains with rapid turnover and diffusion of individual receptors.
\end{abstract}


\maketitle

How the physiological stability necessary for memory storage can be achieved in the presence of rapid molecular turnover and diffusion is a central
problem in neurobiology \cite{crick84}. Synapses, in particular, are believed to be the physiological seat of memory, and rely on the stability of postsynaptic domains containing neurotransmitter receptor molecules, as well as scaffold and a number of other molecules, over days, months, or even longer periods of time \cite{tracht02,grutz02}. Yet, recent experiments have demonstrated that individual receptor \cite{choquet03,triller05,triller08} and scaffold \cite{gray06,specht08,calamai09} molecules leave and enter postsynaptic domains on typical timescales as short as minutes. How can these seemingly contradictory observations---long-term stability and a well-defined characteristic
size of postsynaptic domains on the one hand, rapid molecular turnover and
diffusion on the other hand---be integrated in a unified understanding of postsynaptic domain formation and stability?

Classically, it has been assumed that interactions between presynaptic and postsynaptic neurons play a paramount role in the stability and in setting the characteristic size of synaptic receptor domains \cite{mcallister07}.
Over recent years, though, a number of studies \cite{choquet03,triller08,kummer06}, carried out on a variety of chemical synapses, have indicated that molecular
domains containing synaptic receptor molecules may form spontaneously even in the absence of presynaptic neurons. However, a detailed molecular understanding of the mechanism governing the formation and stability of synaptic receptor domains has remained elusive. In this Letter,
we first discuss a minimal experimental system which enables us to determine the molecular components essential for the self-organization
of stable receptor domains of the characteristic size observed in neurons \cite{meier00,hanus06,calamai09}. On this basis, we then formulate a mathematical model of the formation and stability
of synaptic receptor domains which quantitatively accounts for our experimental observations, and also makes further predictions pertaining to the stability and regulation of synaptic receptor domains.

\begin{figure}[!]
\center
\includegraphics[width=8.5cm]{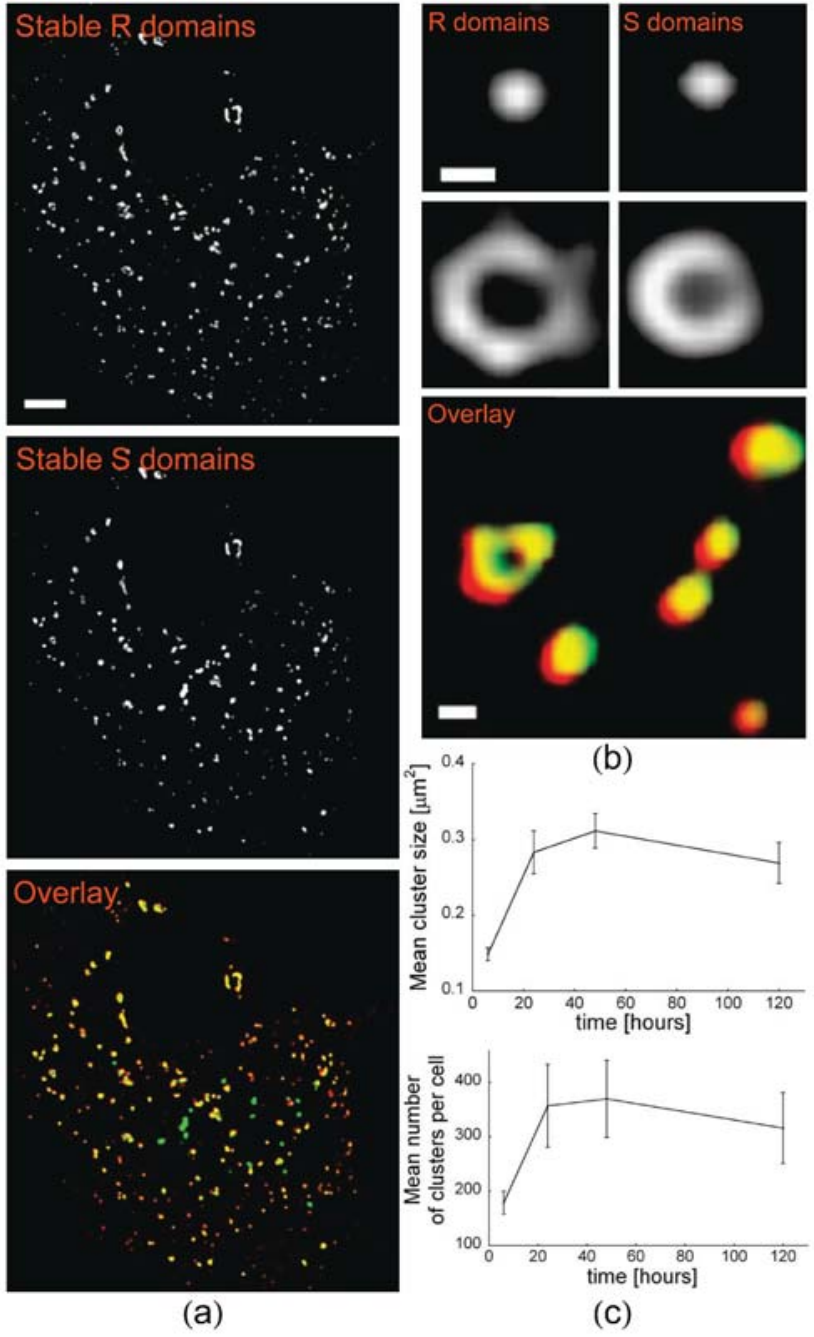}
\caption{\label{fig1}(color online). Experimental results on the formation and stable characteristic
size of RSDs composed
of glycine receptors and gephyrin scaffolds \cite{SOM}. (a) Example of a transfected COS-7 cell with domains on its membrane: Receptor (R, red)
and scaffold (S, green) concentration and overlay (co-domains in yellow). A fraction of the apparent green-labeled scaffolds is endoplasmic. Scale bar, 5 $\mu$m. (b) Examples of selected RSDs at higher resolution. For ease of visualization, the concentration maps of the two molecular species were slightly shifted with respect to one another in the color panel. Scale bars, 0.5 $\mu$m. (c) Mean RSD (cluster) area and number of RSDs (clusters) per cell versus time. Error bars: standard errors; $n>10$ cells from two independent experiments for each point.}
\end{figure}

In our experiments we used single fibroblast cells, which are devoid of the molecular machinery commonly associated with postsynaptic domain formation \cite{mcallister07} but allow for the rapid turnover and diffusion
of receptors observed in neurons \cite{choquet03,triller08}, as well as for interaction of receptors with scaffold molecules. Fibroblast cells were transfected \cite{SOM} with glycine receptors, one of the main receptor types at inhibitory synapses, and their associated scaffolds, gephyrin molecules \cite{calamai09}. In our minimal system, the mere presence of both receptor and scaffold molecules led to the spontaneous emergence of stable receptor-scaffold domains (RSDs) [see Fig. \ref{fig1}(a,b)]. These domains corresponded to a joint enhancement of the receptor and scaffold molecule densities, over a characteristic area of 0.2 to 0.3~$\mu$m$^2$ [Fig. \ref{fig1}(c)]. Once the RSDs were formed, their mean area remained stable over a time scale of days, with little cell-to-cell variability in the mean area of RSDs but larger variability in the mean number of RSDs per cell [Fig. \ref{fig1}(c)]. If only receptors were transfected, in the absence of scaffold molecules, receptor domains did not emerge, apart from possible occurrences of transient microdomains \cite{meier00,meier01}. If only scaffold molecules were transfected, in the absence of receptors, then these formed large intracellular blobs but no association with the cell membrane was detected \cite{meier00,kirsch95}.

The experiments carried out on our minimal system indicate that receptor-scaffold interactions, together with the diffusion properties of each molecular species at the membrane, are \textit{necessary and sufficient} for RSD formation and stability. In particular, the presence of a presynaptic terminal is not essential for the occurrence of stable RSDs. In agreement with previous studies \cite{choquet03,triller08,kummer06}, our results point to a picture in which postsynaptic domains form in the absence of presynaptic stimulation, which subsequently intervenes in their maturation and regulation. Both the characteristic size and the stability of the RSDs observed in our experiments are similar
to those of synaptic receptor domains in neurons. Indeed, when scaffold molecules are transfected to young neurons devoid of synapses, domains of a comparable size arise \cite{meier00}. When they are transfected to mature neurons with synapses, the domain size remains unchanged \cite{hanus06}. Finally, the diffusion properties of receptors are similar in cells with transfected \cite{calamai09} and endogenous \cite{dahan03} scaffold molecules. Thus, we expect that receptors and scaffolds in neurons exhibit the necessary and sufficient properties for RSD formation and stability, as they do in our experiments.

We now turn to the mathematical description of our minimal experimental system. The concentration of receptors is represented by the function $r(x,y,t)$ and that of scaffolds by the function $s(x,y,t)$, where the variables $x$ and $y$ denote coordinates along the cell membrane, and the variable $t$ denotes time. The spatiotemporal evolution of these fields is governed by the reaction-diffusion equations
\begin{eqnarray} \label{cont1}
\frac{\partial r}{\partial t}&=& F(r,s)+\nu_r \nabla \left[(1-s)\nabla r+r \nabla s\right]\,,\\ \label{cont2}
\frac{\partial s}{\partial t}&=& G(r,s)
+\nu_s \nabla \left[(1-r)\nabla s+s \nabla r\right]\,,\label{cont2}
\end{eqnarray} 
where $F$  and $G$ are simple (cubic) polynomials in $r$ and $s$ that describe the reactions in our system \cite{SOM}, and $\nu_r$  and $\nu_s$ are the receptor and scaffold diffusion coefficients. The nonlinear corrections to the standard diffusion terms $\nu_r \nabla^2 r$  and $\nu_s \nabla^2 s$  in Eqs.~(\ref{cont1}) and~(\ref{cont2})
arise from the constraint $0\leq r+s \leq1$, where we have normalized $r$ and $s$ so that the maximum concentration of receptors and scaffolds is equal to $1$, and account for steric repulsion \cite{choquet03,triller08} of receptors and scaffolds
in the confined membrane environment of a living cell. Experimental studies
\cite{choquet03,triller08,gray06,specht08,triller05,calamai09,meier01}
of the diffusion properties of glycine receptors and gephyrin scaffolds,
as well as of other types of synaptic receptors and scaffolds, yield
$\nu_r>\nu_s$.

The reaction and diffusion properties of receptors and scaffolds 
\cite{choquet03,triller08,gray06,calamai09,triller05,specht08,meier01} suggest that Eqs.~(\ref{cont1}) and~(\ref{cont2}) exhibit pattern formation via a Turing instability \cite{turing52,cross93} which emerges from the interplay between the two molecular species in our system: Receptors diffuse quickly and tend to repel nearby molecules, whereas scaffolds diffuse more slowly and tend to attract nearby molecules. In agreement with experiments, the formation of synaptic receptor domains via a Turing mechanism necessarily relies on the presence of both receptors and scaffolds. Expressions of the reaction terms $F$ and $G$ in Eqs.~(\ref{cont1}) and~(\ref{cont2}) are obtained from the relevant chemical interactions, reported previously 
\cite{choquet03,triller08,gray06,calamai09,triller05,specht08}, together
with the general mathematical constraints associated with Turing
instabilities \cite{cross93}, a point we return to below. Reaction-diffusion
models akin to the one described here have, in recent years, been used to describe molecular localization during cell division \cite{howard01,kulkarni04,loose08}, and are to be contrasted with models of domain formation which rely on phase separation and coarsening \cite{gamba07}.

\begin{figure}[t!]
\center
\includegraphics[width=8.5cm]{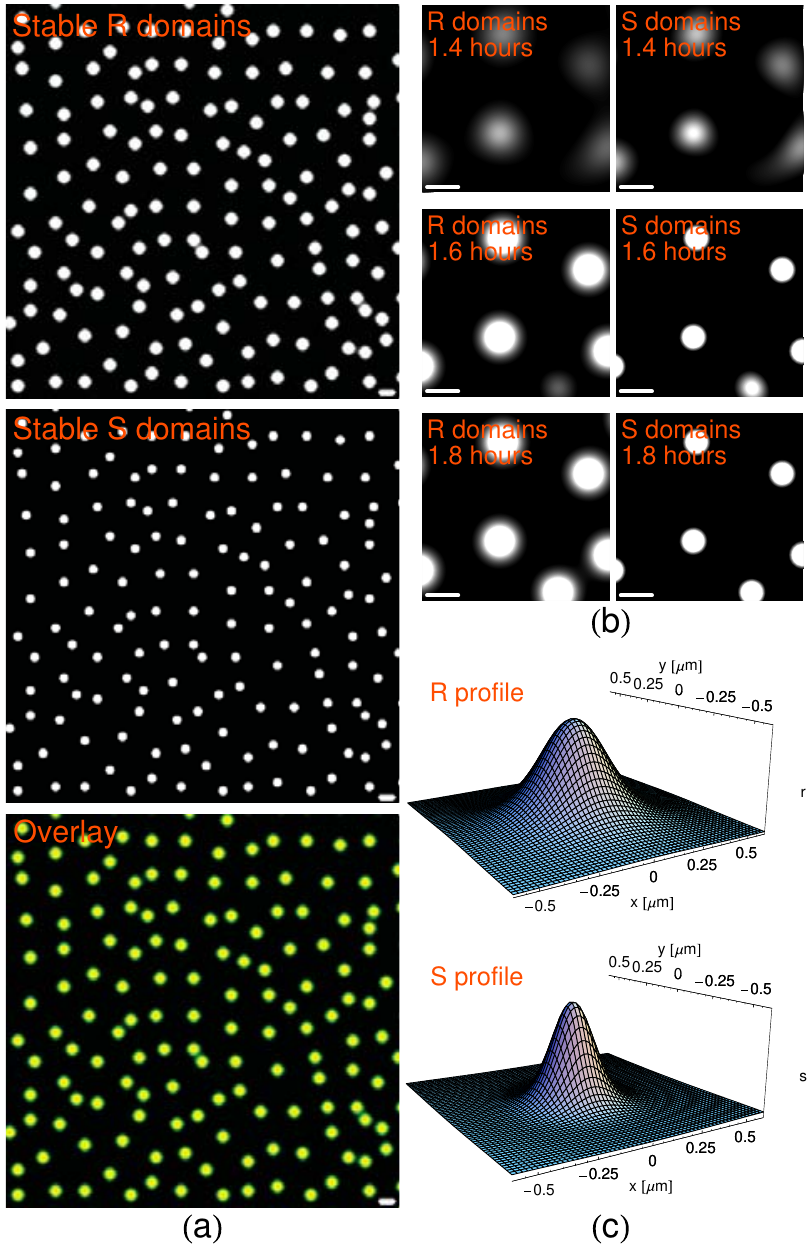}
\caption{\label{fig2}(color online). Model results on the formation and stable characteristic size of RSDs \cite{SOM}. (a) Irregular patterns of stable RSDs, with an area of approximately 0.2 to 0.3~$\mu$m$^2$ each, emerge on a timescale of hours. The distributions of receptors (upper panel) and scaffolds (middle panel) are overlayed (lower panel), and the domain shapes and patterns are stable. (b) Formation and (c) shape of RSDs at higher resolution. Scale bars, 0.5~$\mu$m.}
\end{figure}

We simulated Eqs.~(\ref{cont1}) and~(\ref{cont2}), starting from random initial conditions, with units of space and time set by the diffusion coefficient of unbound receptor molecules and the rate of receptor endocytosis. Using typical values for these parameters taken from experiments 
\cite{choquet03,triller08,gray06,calamai09,triller05,specht08,meier01}, we found that irregular patterns of stable RSDs similar to experimental ones emerged over a timescale of hours [see Figs.~\ref{fig2}(a) and~\ref{fig1}(a)]. Individual domains resulted from a coordinated increase of receptor and scaffold densities. Occasionally, we observed sets of closely-spaced RSDs in the outcomes of simulations, resulting from initial random fluctuations [see Figs.~\ref{fig2}(a,b)]. At lower resolution, these appeared as larger than average, irregularly shaped domains reminiscent of similar instances obtained experimentally [see the lower panels in Fig.~\ref{fig1}(b)]. Inclusion of molecular noise in our
reaction-diffusion model is expected to further distort the simulated patterns without changing their overall characteristics \cite{butler09}. Moreover, we observed in our simulations that receptor aggregation trails behind scaffold aggregation in time [Fig.~\ref{fig2}(b)], which is in fact a general feature of our model and is also in agreement with experimental observations~\cite{kirsch93,bechade96}.

The Turing mechanism implies that RSDs attain a steady state once a dynamical equilibrium between strong receptor diffusion and the attractive effect of scaffolds is reached. In our simulations of Eqs.~(\ref{cont1}) and~(\ref{cont2}) we found that, for the diffusion coefficient of glycine receptors reported in experiments \cite{choquet03,meier01,triller05,calamai09,triller08}, the resulting characteristic size of RSDs was comparable to that obtained in experiments [see Figs.~\ref{fig2}(b,c) and~\ref{fig1}(c)]. This result necessitated only gross adjustment of the reaction rates, so that Eqs.~(\ref{cont1}) and~(\ref{cont2}) would exhibit a Turing instability but, apart from that, the characteristic size of domains was found to be largely insensitive to variations in reaction rates or, indeed, to the reaction kinetics considered. The timescale for the formation of stable RSDs in our simulations [see Fig.~\ref{fig2}(b)] was set by the rate of receptor endocytosis. For the measured range of values of the rate of receptor endocytosis \cite{choquet03,triller08,specht08} this timescale was also in broad agreement with experimental observations [see Fig.~\ref{fig1}(c)]. Further consistency checks between Eqs.~(\ref{cont1}) and~(\ref{cont2}) and experimental results would pertain to the geometry of domain edges, the effect of changes in diffusion rates on domain size and stability, and the relative extent of the enhanced concentrations of receptors and scaffolds in RSDs. In particular, our model predicts that once dynamical equilibrium has set in, the receptor profile is wider than the scaffold profile in any given domain
[see Fig.~\ref{fig2}(c)].

While, in accordance with experimental observations
\cite{choquet03,triller08,gray06,calamai09,triller05,specht08,meier01}, we allowed \cite{SOM} for a variety of interactions between receptors
and scaffolds when simulating Eqs.~(\ref{cont1}) and~(\ref{cont2}), we found that only a handful of chemical reactions were crucial for the formation and stability of RSDs via a Turing instability. To lowest order, these reactions correspond to $R \to R_b$ and $R_b+S \to R+S$ for the receptors, and to $S \to S_b$ and $S_b + 2S \to 3S$ for the scaffolds, respectively. In these
expressions, the symbols $R$ and $S$ stand for receptors and scaffolds at the membrane, and $R_b$ and $S_b$ denote molecules in the bulk of the cell. In particular, the reaction $S_b + 2S \to 3S$, in which a scaffold molecule from the bulk is adsorbed onto the membrane by two other scaffold molecules into a trimer, is key to domain formation, whereas the simpler reaction $S_b + S \to 2S$  alone is not sufficient. Indeed, gephyrin scaffold molecules are thought to form both dimers and trimers under the usual conditions in which neural domains are observed \cite{choquet03}. However, if trimerization is prevented, no domains (or only very small ones) appear \cite{calamai09}.

The above results demonstrate how stable synaptic receptor domains can emerge in the absence of presynaptic stimulation. In a synapse, however, presynaptic activity regulates \cite{shepherd07} the concentration of receptors
in the postsynaptic domain. Our reaction-diffusion
model suggests novel mechanisms for how such regulation may be achieved.
It has been observed \cite{specht08,bannai09} that the diffusion of receptors on the postsynaptic membrane can be modified through binding of presynaptic neurotransmitters. Similarly, scaffold diffusion may be \cite{chih05,dean06} modulated by synaptic activity. This suggests that local modification of the diffusion properties of receptors or scaffolds may contribute to the regulation of postsynaptic domains. As a simple phenomenological perturbation to our model, we therefore implemented pre- and postsynaptic interactions through a local increase in the receptor diffusion rate which, within the framework of our reaction-diffusion model, has the same effect as a local
decrease in the scaffold diffusion rate.

\begin{figure}[t!]
\center
\includegraphics[width=8.5cm]{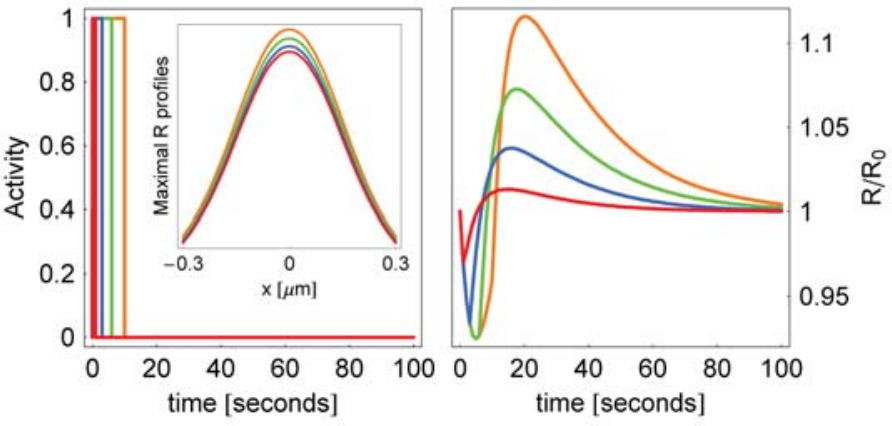}
\caption{\label{fig4}(color online). Model results on the regulation of mature RSDs \cite{SOM}. Left panel: temporal profiles of step stimulations. Inset: receptor concentration profiles at times of maximum domain size. Right panel: time course of the in-domain receptor population size, $R$, following stimulation, normalized by the in-domain receptor population size in the absence of any stimulation, $R_0$.}
\end{figure}

As shown in Fig.~\ref{fig4}, our model predicts that modulation of receptor or scaffold diffusion yields a transient increase of the in-domain receptor population following presynaptic activity. Clearly, this speculative, purely biophysical mechanism may coexist with biochemical mechanisms of postsynaptic plasticity. Applying a few seconds of stimulation at a time (Fig.~\ref{fig4}, left panel), we found a correlation between the increase in in-domain receptor population and the duration of stimulation (Fig.~\ref{fig4}, right panel). After an initial, short-lived suppression, the population increase lasted for a few tens of seconds---the timescale typically associated with short-term
plasticity. In this window of time, RSDs were richer in receptors (Fig.~\ref{fig4}, inset) and, hence, yielded a larger synaptic efficacy. This phenomenon has a simple explanation in terms of the Turing instability exhibited by our model: Enhanced (diminished) receptor (scaffold) diffusion depletes the receptor (increases the scaffold) population in a transient manner which, because receptors are repulsive and scaffolds are attractive, in turn attracts even more receptors and scaffolds into RSDs.

In summary, we have used a minimal experimental system devoid of synaptic machinery to show that neurotransmitter receptor domains of the stable characteristic size observed in neurons can emerge from nothing more than interactions between receptors and scaffolds, together with the rapid diffusion of receptors on the cell membrane. A reaction-diffusion model quantitatively accounts for our experimental results, yielding spontaneous formation of stable receptor domains and their observed characteristic size, as well as new putative mechanisms for the regulation of synaptic strength. Collectively, our results show how stable synaptic receptor domains may form even in the absence of presynaptic stimulation \cite{choquet03,triller08,kummer06}, and how rapid turnover and
diffusion of receptors \cite{choquet03,triller08}, far from being a hindrance, may in fact be crucial \cite{crick84} for ensuring overall stability and delicate control of synaptic receptor domains.

This work was supported by the Austrian Science Fund (C.A.H.), the Pierre-Gilles de Gennes Foundation and a FEBS grant (M.C.), the NSF through grant No. DMR-08-03315 and a Visiting Professorship at the Ecole Normale Sup\'erieure (M.K.), the Inserm UR 789 (A.T.), and the CNRS through UMR 8550 (R.A.d.S.).


\begin{thebibliography}{99}

\bibitem{crick84} F. Crick, Nature \textbf{312}, 101 (1984).

\bibitem{tracht02} J. T. Trachtenberg \textit{et al.}, Nature \textbf{420}, 788 (2002).

\bibitem{grutz02} J. Grutzendler, N. Kasthuri, and W.-B. Gan, Nature \textbf{420}, 812 (2002).

\bibitem{choquet03} D. Choquet and A. Triller, Nat. Rev. Neurosci. \textbf{4}, 251 (2003).

\bibitem{triller05} A. Triller and D. Choquet, Trends Neurosci. \textbf{28}, 133 (2005).

\bibitem{triller08} A. Triller and D. Choquet, Neuron \textbf{59}, 359 (2008).

\bibitem{gray06} N. W. Gray \textit{et al.}, PLoS Biology \textbf{4}, 2065 (2006).

\bibitem{specht08} C. G. Specht and A. Triller, Bioessays \textbf{30}, 1062
(2008).  

\bibitem{calamai09} M. Calamai \textit{et al.}, J. Neurosci. \textbf{29}, 7639 (2009).

\bibitem{mcallister07} A. K. McAllister, Ann. Rev. Neurosci. \textbf{30}, 425 (2007).

\bibitem{kummer06} T. T. Kummer, T. Misgeld, and J. R. Sanes, Curr. Opin. Neurobiol. \textbf{16}, 74 (2006); and references therein.

\bibitem{meier00} J. Meier \textit{et al.}, J. Cell Sci. \textbf{113}, 2783
(2000).

\bibitem{hanus06} C. Hanus, M.-V. Ehrensperger, and A. Triller, J. Neurosci. \textbf{26}, 4586 (2006). 

\bibitem{SOM} For further details, see the supplementary material posted with this article.

\bibitem{meier01} J. Meier \textit{et al.}, Nat. Neurosci. \textbf{4}, 253
(2001).

\bibitem{kirsch95} J. Kirsch, J. Kuhse, and H. Betz, Mol. Cell. Neurosci. \textbf{6}, 450 (1995). 

\bibitem{dahan03} M. Dahan \textit{et al.}, Science \textbf{302}, 442 (2003).

\bibitem{turing52} A. M. Turing, Phil. Trans. B \textbf{237}, 37 (1952).

\bibitem{cross93} M. C. Cross and P. C. Hohenberg, Rev. Mod. Phys. \textbf{65}, 851 (1993).

\bibitem{howard01} M. Howard, A. D. Rutenberg, and S. de Vet, Phys. Rev. Lett. \textbf{87}, 278102 (2001).

\bibitem{kulkarni04} R. V. Kulkarni \textit{et al.}, Phys. Rev. Lett. \textbf{93},
228103 (2004).

\bibitem{loose08} M. Loose \textit{et al.}, Science \textbf{320}, 789 (2008).

\bibitem{gamba07} A. Gamba \textit{et al.}, Phys. Rev. Lett. \textbf{99},
158101 (2007).

\bibitem{butler09} T. Butler and N. Goldenfeld, Phys. Rev. E \textbf{80},
030902(R) (2009).

\bibitem{kirsch93} J. Kirsch \textit{et al.}, Nature \textbf{366}, 745 (1993).

\bibitem{bechade96} C. B\'echade \textit{et al.}, Eur. J. Neurosci. \textbf{8}, 429 (1996). 

\bibitem{shepherd07} J. D. Shepherd and R. L. Huganir, Ann. Rev. Cell Dev. Biol. \textbf{23} 613 (2007).

\bibitem{bannai09} H. Bannai \textit{et al.}, Neuron \textbf{62}, 670 (2009).

\bibitem{chih05} B. Chih, H. Engelman, and P. Scheiffele, Science \textbf{307}, 1324 (2005).

\bibitem{dean06} C. Dean and T. Dresbach, Trends Neurosci. \textbf{29}, 21
(2006).


\end{thebibliography}
\end{document}


\begin{center}

\noindent
{\huge Formation and Stability of Synaptic Receptor Domains:}

\smallskip

{\huge Supplementary Material}

\bigskip

Christoph A. Haselwandter, Martino Calamai, Mehran Kardar, Antoine Triller,\\
and Rava Azeredo da Silveira

\end{center}

\noindent
A summary of experimental materials and methods can be found in Sec.~\ref{secExpMeth},
while Secs.~\ref{secsimulations} and~\ref{secPlasticity} provide supplementary
information pertaining to the computer simulations described in the main
text.

\section{Experimental materials and methods}
\label{secExpMeth}

\begin{figure}[t]
\center
\includegraphics[width=7.6cm]{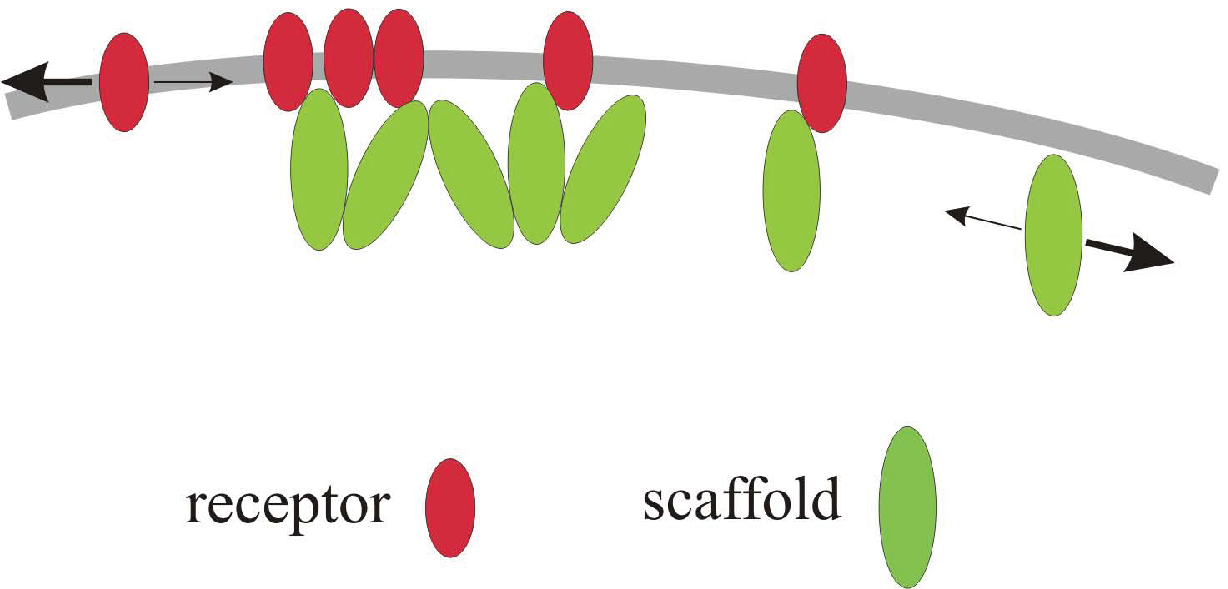}
\caption{\label{illust}Schematic view of glycine receptors (red) and gephyrin scaffold proteins (green). Both molecular species can diffuse (arrows) and binding can exist between glycine receptors and gephyrin and between gephyrin and gephyrin. Collectively, glycine receptors and gephyrin scaffolds can form domains.}
\end{figure}

As illustrated in Fig.~\ref{illust}, our minimal experimental system allows
for interactions between glycine receptors and gephyrin scaffold proteins, and the diffusion of these molecules at the cell membrane. The construction of the chimeric cDNA for the GlyR$\alpha$1 subunit bearing the gephyrin binding sequence ($\beta$gb) and of the venus-tagged gephyrin (VeGe) have been described earlier \cite{meier00}. Subconfluent African green monkey kidney (COS-7) cells were co-transfected (FuGENE 6, Roche Applied Science, France) using a total of 2$\mu$g plasmid DNA at a GlyR gephyrin stoichiometry of 2:1. At successive times [as indicated in Fig.~1(c) of the main text], GlyR immunolabelling was carried out with primary antibody and secondary antibody coupled to Cy3 (as described previously for the visualization of cell surface receptors \cite{rosenberg01}), and VeGe fluorescence was visualized directly. The area of GlyR-Vege clusters was quantified using a multidimensional image analysis as described previously \cite{calamai09}.

\section{Stable patterns of synaptic receptor domains}
\label{secsimulations}

In accordance with experimental observations
\cite{choquet03a,triller08,okabe99,gray06,calamai09,triller05,specht08}, we allowed for a variety of chemical reactions between receptors
and scaffolds when simulating Eqs.~(1) and~(2) of the main text,
\begin{eqnarray} \nonumber
F(r,s)&=&-b\left(r-\frac{s}{\bar s}\frac{1-r-s}{1-\bar r-\bar s}\bar r\right)
-m_1 \frac{1-r-s}{1-\bar r-\bar s}\left(r-\bar r\right)
+m_2\frac{1-r-s}{1-\bar r-\bar s}\frac{r}{\bar r}\left(s-\bar s\right)
\\\label{F3l2}
&=&\underbrace{-b r}_{R\rightarrow R_b}
\underbrace{+m_1
\frac{1-r-s}{1-\bar r-\bar s}\bar r}_{R_b\rightarrow R}
\underbrace{+b\frac{1-r-s}{1-\bar r-\bar s} \frac{\bar r}{\bar s} s}_{R_b+S\rightarrow R+S} \underbrace{-\left(m_1 +m_2\frac{\bar s}{\bar r}\right) \frac{1-r-s}{1-\bar r-\bar s} r}_{M_b+R\rightarrow R_b+M_b}\underbrace{+\frac{m_2}{\bar r} \frac{1-r-s}{1-\bar r-\bar s}r s}_{R_b+R+S\rightarrow2
R+S}\,,
\\ \nonumber
G(r,s)&=&-\beta\left(s-\frac{1-r-s}{1-\bar r-\bar s}\bar s\right)+\mu \frac{1-r-s}{1-\bar r-\bar s} \frac{s}{\bar s}\left(s-\bar s\right)\\ \label{G3l2}
&=& \underbrace{-\beta s}_{S\rightarrow S_b}\underbrace{+\beta
\frac{1-r-s}{1-\bar r-\bar s} \, \bar s}_{S_b\rightarrow S}
\underbrace{-\mu \frac{1-r-s}{1-\bar r-\bar s} s}_{M_b+S\rightarrow
S_b+M_b} \underbrace{+\frac{\mu}{\bar s} \frac{1-r-s}{1-\bar r-\bar s}s^2}_{S_b+2S\rightarrow3S}\,,
\end{eqnarray}
where we have indicated below each monomial term the corresponding chemical reaction that it describes. The chemical reactions are expressed in terms of $R$ and $S$, which stand for receptors and scaffolds at the membrane, and $R_b$, $S_b$, and $M_b$, which denote molecules in the bulk of the cell. The expressions in Eqs.~(\ref{F3l2}) and~(\ref{G3l2}) are arranged so that the reaction terms exhibit a stable homogeneous fixed point at the concentrations $(r,s)=(\bar
r,\bar s)$ since $F(\bar r,\bar s)=0$  and $G(\bar r,\bar s)=0$. It then follows from Eqs.~(1) and~(2) of the main text that the receptor and scaffold densities are stationary in time if $(r,s)=(\bar r,\bar s)$ everywhere on the membrane. Random perturbations of this fixed point trigger pattern formation
via a Turing instability \cite{turing52,cross93,meinhardt82a,walgraef97,epstein98}.

To simulate Eqs. (1) and (2) of the main text it is convenient to introduce the dimensionless variables
\begin{equation} \label{transdim}
x\to \widetilde{x}=\left(\frac{b}{\nu_r}\right)^{1/2}\,x\,,\quad y\to \widetilde{y}=\left(\frac{b}{\nu_r}\right)^{1/2}\,y\,, \quad t\to \widetilde{t}=b\,t\,,
\end{equation}
in terms of which Eqs. (1) and (2) of the main text become
\begin{eqnarray}  \label{cont1nod}
\frac{\partial r}{\partial \widetilde t}&=& \widetilde F(r,s)+\widetilde \nabla \left[(1-s)\widetilde \nabla r+r \widetilde \nabla s\right]\,,\\
\frac{\partial s}{\partial \widetilde t}&=& \widetilde G(r,s)
+\widetilde \nu_s \widetilde \nabla \left[(1-r)\widetilde \nabla s+s \widetilde \nabla r\right]\,,\label{cont2nod}
\end{eqnarray}
where $\widetilde F(r,s)=\widetilde F\left(r,s;\widetilde{m_1},\widetilde{m_2}\right)$
and $\widetilde G(r,s)=\widetilde G\left(r,s;\widetilde{\beta},\widetilde{\mu}\right)$,
with the parameters $\left(\widetilde{m_1},\widetilde{m_2},\widetilde{\beta},\widetilde{\mu}\right)
=\frac{1}{b} \left(m_1,m_2,\beta,\mu\right)$, $\widetilde{\nu_s}=\nu_s/\nu_r$, and $\widetilde{\nabla}=\left(\frac{\partial}{\partial \widetilde{x}},\frac{\partial}{\partial \widetilde{y}}\right)$. We restore physical dimensions to the solutions of Eqs.~(\ref{cont1nod}) and~(\ref{cont2nod}) using \cite{choquet03a,meier01a,triller05,calamai09,triller08,specht08}
the value $b=10^{-1}$~sec$^{-1}$ which corresponds
to the rate
of endocytosis of individual receptor molecules, and the value $\nu_r=10^{-2}$~$\mu$m$^2$~sec$^{-1}$
for the diffusion coefficient of free receptor molecules. Smaller values
\cite{choquet03a,triller08,specht08} for the rate of receptor endocytosis produced very similar patterns as in Fig.~2 of the main text, but with a slower temporal evolution towards the steady state.

Equations~(\ref{cont1nod}) and~(\ref{cont2nod}) were simulated on a square lattice with side length 6.3 $\mu$m using periodic boundary conditions and a spatial grid with spacing 0.063 $\mu$m. We used initial conditions for $r$ and $s$ which were randomly distributed in the interval $[0,0.01]$, well below our choices $(\bar r, \bar s)=(0.05,0.05)$ for the numerical values of $\bar r$ and $\bar s$. Simulations with a smaller grid spacing and other choices for the maximum amplitude of initial perturbations produced results similar to those presented in Fig.~2 of the main text. Moreover, since Eqs.~(\ref{cont1nod}) and (\ref{cont2nod}) can generally admit several homogeneous fixed points in addition to $(\bar r,\bar s)$, we checked that, for a given set of dimensionless parameters, $r$ and $s$ indeed approached the fixed point $(\bar r,\bar s)$ for homogeneous initial conditions in the range $(0,1)$. For the simulations discussed in
the main text, we used the parameter values $(\widetilde m_1, \widetilde m_2, \widetilde \beta, \widetilde \mu, \widetilde \nu_s)=\left(0.4,10,0.5,0.7,0.02\right)$.

\section{Regulation of synaptic receptor domains}
\label{secPlasticity}

Accounting for local variations of receptor and scaffold diffusion rates
in our reaction-diffusion model, we obtain the following
modified versions of Eqs.~(1) and (2) of the main text:
\begin{eqnarray} \label{cont1plas}
\frac{\partial r}{\partial t}&=& F(r,s)+\nu_r \nabla \left[(1-s)\nabla (
D_r r)+D_r r \nabla s-r^2 \nabla D_r\right]\,,\\
\frac{\partial s}{\partial t}&=& G(r,s)
+\nu_s \nabla \left[(1-r)\nabla (
D_s s)+D_s s \nabla r-s^2 \nabla D_s\right]\,,\label{cont2plas}
\end{eqnarray}
where the functions $D_r=D_r(x,y,t)$ and $D_s=D_s(x,y,t)$ represent spatial
and temporal variations in the receptor and scaffold diffusion rates, and
Eqs.~(1) and (2) of the main text are recovered for $D_r(x,y,t)=D_s(x,y,t)=1$.

The results shown in Fig.~3 of the main text were obtained from simulations
of the dimensionless versions of Eqs.~(\ref{cont1plas}) and~(\ref{cont2plas}),
\begin{eqnarray} \label{cont1plasnod}
\frac{\partial r}{\partial \widetilde t}&=& \widetilde F(r,s)+ \widetilde
\nabla \left[(1-s)\widetilde \nabla (
D_r r)+D_r r \widetilde \nabla s-r^2 \widetilde \nabla D_r\right]\,,\\
\frac{\partial s}{\partial \widetilde t}&=& \widetilde G(r,s)
+\widetilde \nu_s \widetilde \nabla \left[(1-r)\widetilde \nabla (
D_s s)+D_s s \widetilde \nabla r-s^2 \widetilde \nabla D_s\right]\,,\label{cont2plasnod}
\end{eqnarray}
where we use the same notation as in Sec.~\ref{secsimulations}. We set $D_s(x,y,t)=1$
and took the function $D_{r}(x,y,t)$ to be a sum of Gaussians in the spatial variables with a threshold dependence on time:
\begin{equation} \label{defDr}
D_{r}\left(\widetilde x,\widetilde y,\widetilde t\right)=1 + \widetilde
A_{r} \sum_{i,j,k} \theta\left(\widetilde t-\widetilde t_i\right)\exp\left[-\frac{(\widetilde
x-\widetilde x_j)^2+(\widetilde y-\widetilde y_k)^2}{\widetilde l_{r}}\right]\,,
\end{equation}
where the step function $\theta\left(\widetilde t\right)$ is defined by
\begin{equation} \label{distheta}
\theta\left(\widetilde t \right)=\begin{cases}1 & \text{if $ \widetilde
t\ge0$,} \\ 0 & \text{if
$ \widetilde t<0$}.
\end{cases}
\end{equation}
We set $\widetilde \ell_r=3$ and $\left|\widetilde A_{r}\right|=1/5$ for all simulations shown in Fig.~3 of the main text, and the periods of activity lasted 1~sec, 3~sec, 6~sec, and 10~sec.